\documentstyle[graphicx]{lamuphys}
\makeatletter
\let\chapter\hid@chapter
\makeatother
\begin{document}
\pagenumbering{arabic}
\title{Ginzburg-Landau Theory for Impure Superfluid $^3$He}

\author{E.V. Thuneberg}

\institute{Low Temperature Laboratory, Helsinki University 
of 
Technology, P.O. Box 2200, 02015 TKK, Finland}

\maketitle

\begin{abstract}
Liquid $^3$He at low temperatures is an ideal substance to study because of its
natural purity. The superfluid state, which appears at temperatures below 3
mK, has many unusual and exciting properties. These have extensively
been studied experimentally, and the results are in many cases well understood
theoretically using quasiclassical approximation. In this work we study the
case that liquid $^3$He is inside of aerogel. Aerogel is a very porous
substance, where 98\% of the volume can be empty. The purpose is to understand
how the properties of the superfluid are modified when the quasiparticles are
scattered from an impurity, such as aerogel. We study extensively the {\it
homogeneous scattering model}. From it we derive the Ginzburg-Landau theory of
impure superfluid $^3$He. We give expressions for measurable quantities and
compare them with experiments. We consider random anisotropy and its effect on
the NMR properties. More sophisticated scattering models are briefly
discussed.\footnote{Work done in collaboration with M. Fogelstr\"om, S.K. Yip,
J.A. Sauls, R. H\"anninen, and T. Set\"al\"a} \end{abstract}
\section{Introduction}

Liquid $^3$He at low temperatures is the purest substance in Nature because
usual impurities simply will fall down by gravity. This is one of the reasons why
liquid $^3$He, in spite of its strong particle-particle interactions, is one of
the best understood condensed matter systems (\cite{3Hereview}). Especially the
superfluid state, which occurs at temperatures below 3 mK, shows many
complicated but still well understood phenomena. But the purity can also be a
limitation because often important physical effects exist only because of
impurities. For example, except of a few marginal cases, all pure elemental
superconductors are of type I. That is, type II superconductivity, where magnetic
field penetrates into the superconductor as quantized flux lines, exist because
of impurities which radically change the properties of pure elements. 

\begin{figure}[bt]
\begin{center}\leavevmode
\includegraphics[width=0.5\linewidth]{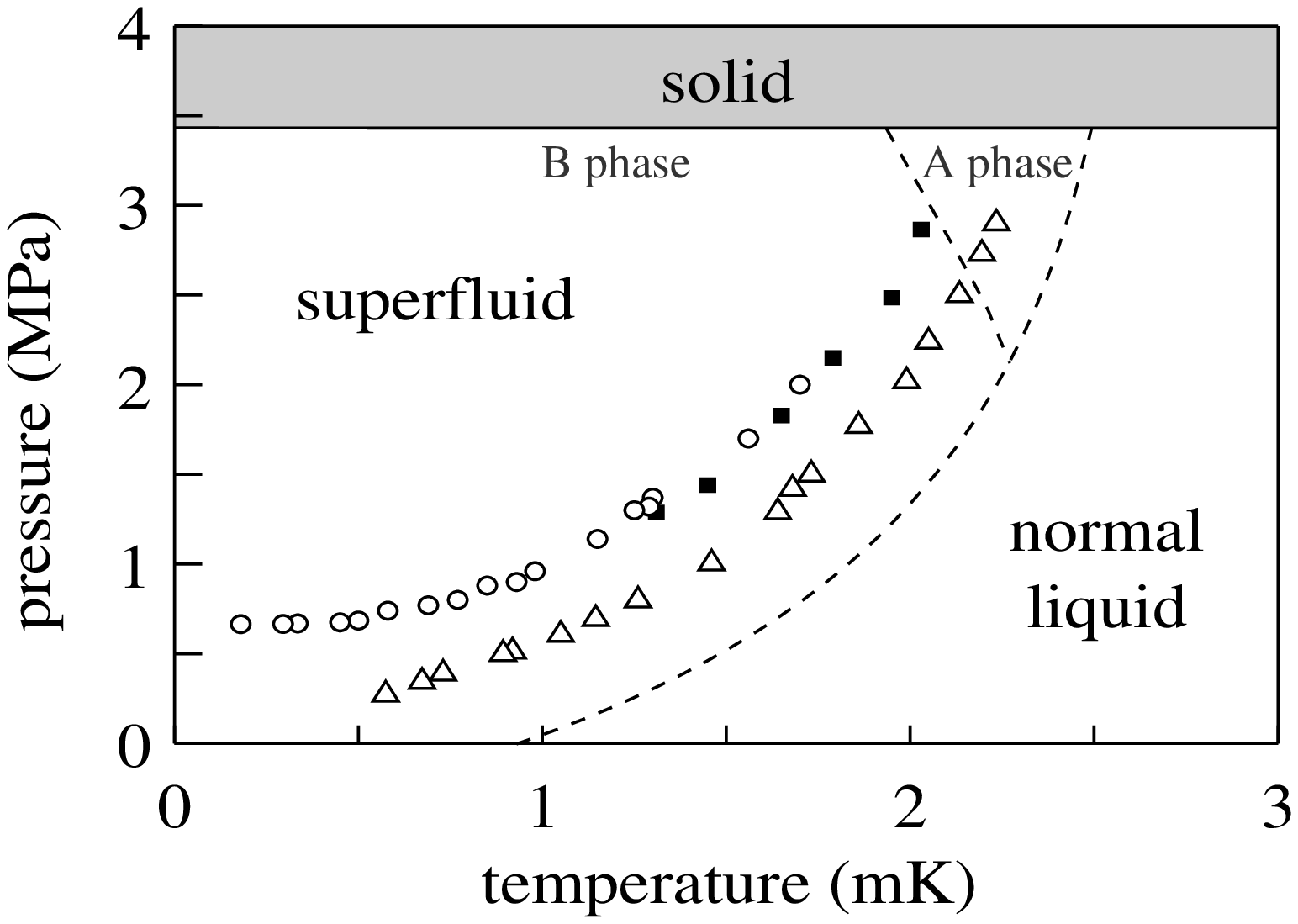}
\bigskip
\caption[f.tcpres]{ 
The measured superfluid transition points of $^3$He in aerogel: 
Porto and Parpia (1995, \font\apu=cmsy7  {\apu\char'064}),
Sprague et al. (1995, \kern .2ex\vrule height .95ex width .8ex depth -.15ex
\kern .2ex), and Matsumoto et al. (1997, $\circ$). The dashed lines denote
the normal-superfluid and the A-B transition lines in pure $^3$He. The A and B
phases also refer to pure $^3$He. 
}\label{f.tcpres}\end{center}\end{figure}

One type of impurity is formed by the surfaces of the container that hold the
liquid. There has been several studies of liquid $^3$He in different
confined geometries, for example between parallel plates or in packed powders
(for example, \cite{TholenParpia}). In these cases the effect on $^3$He arises
from walls, i.e.\ from 2-dimensional interfaces between $^3$He and a foreign
object. It is very difficult to support zero-dimensional (point-like)
impurities in liquid $^3$He. However, there is the
intermediate case of one-dimensional impurities. This type of impurity can
approximately be realized by silica aerogels. These are very porous materials
where, for example, 98\% of the volume is empty. Several recent experiments
investigate $^3$He in aerogel (\cite{Porto,Matsumoto,Sprague1}, 1996). It is
found that, similar to the bulk liquid, $^3$He in aerogel goes to the
superfluid state, but both the superfluid transition temperature $T_{\rm c}$ and
the amplitude of the superfluid state are reduced compared to the pure case. The
measured suppression of $T_{\rm c}$ is shown in Fig.\ \ref{f.tcpres}.
 
In this article we present some theoretical ideas concerning $^3$He in
aerogel.   We first describe the structure of aerogel (Section
\ref{s.aerogel}). The main effect of aerogel on superfluid $^3$He arises
from scattering of the quasiparticles of $^3$He from the aerogel impurity. This
effect can be calculated to a good approximation using the quasiclassical
theory. The general assumptions of quasiclassical scattering models are
discussed in Section \ref{s.qcgeneral}. In Section \ref{s.hsm} we study a
model where the scattering is considered as homogeneously distributed. We will
limit to temperatures in the neighborhood of the transition temperature
$T_{\rm c}$, and derive the Ginzburg-Landau theory. We calculate measurable
quantities and compare them with experiments. We find that more sophisticated
scattering models are needed in order to explain the measurements
quantitatively (Section \ref{s.iism}). The anisotropy of the scattering is
considered in Section \ref{s.ahsm}. There we study a random field model using
similar arguments as presented by Imry and Ma (1975), and find that the NMR
properties crucially depend on the anisotropy.  
 
\section{Aerogel}\label{s.aerogel}

A nice introduction to aerogels is given by Fricke (1988). The fabrication uses
a gelation process in the liquid phase. A variety of aero{\it gel} materials are
possible, but the present ones are of silica, (SiO$_2$)$_n$. In order to get
{\it aero}gel it is crucial to preserve the gel structure when the solvent is
removed. This is not trivial because the surface tension of the liquid--gas
interface would make the gel to collapse in straightforward drying.
Fortunately, it is possible to go continuously from the liquid to the gas
phase. For that one has to move along such a path that goes
around the critical point
($T_0$, $p_0$) in the temperature--pressure plane. In order to avoid too high
temperatures and pressures, methanol (512 K, 8 MPa) is preferred to water (647
K, 22 MPa) as a solvent. 

The experiments with superfluid $^3$He use samples where the aerogel fills only
2\% of the total volume, $V=0.02$. The surface to volume ratio is determined
by measurements with $^4$He, and they give $A=260 000\ {\rm  cm}^{-1}$
(\cite{KMC}). Based on these numbers alone, we can make some estimations of the
dimensions in aerogel. If we assume that the structure is a lattice of
one-dimensional strands, we calculate the strand diameter ${4V/ A}=3$ nm
and the distance between neighboring strands ${\sqrt{4\pi V}/A}=20$ nm. The
mean free path for straight line trajectories is estimated as $\ell={4/A}=150$
nm. Although this picture is certainly very idealized, the numbers are not very
different from the ones obtained by other means such as transmission electron
microscopy or low-angle x-ray scattering. The distance between strands
certainly has large variation because of fluctuations in the local aerogel
density.

The next step in understanding superfluid $^3$He in aerogel is to study the
characteristic lengths of $^3$He.  The smallest scale is the average distance
between  neighboring $^3$He atoms. This is on the same order of magnitude as
the Fermi wave length $\lambda_{\rm f}= 2\pi/k_{\rm f}\approx 0.7$ nm. This is
somewhat smaller than the diameter of the strands. A much larger scale is the
coherence  length  $\xi_0={\hbar v_{\rm f}/ 2\pi k_{\rm B}T_{\rm c0}}$.  Here
$T_{\rm c0}$ is the superfluid transition  temperature in bulk $^3$He and
$v_{\rm f}$ is the Fermi velocity. As a function of pressure,
$\xi_0$ decreases monotonically from 77 nm (zero pressure) to 16 nm (3.4 MPa,
the solidification pressure). The coherence length is on the same order as the
average strand spacing, but taking into account fluctuations in strand
density, we can well expect voids of size larger than $\xi_0$. Finally, the
mean free path of quasiparticles in pure $^3$He is $1\ \mu$m or more at the
temperatures we are interested in ($T<3$ mK). Thus the scattering from strands
is the dominant factor that limits the mean free path in aerogel.

\section{Quasiclassical Scattering Models}\label{s.qcgeneral}

Before going into the details of different scattering models, it is useful to
study the general assumptions. Large part of the ideas presented here are given
in a more mathematical form by Buchholtz and Rainer (1979). The following is
applicable to both metals and
$^3$He at low temperatures. Here a ``particle'' refers to an electron in the
former case and to a $^3$He atom in the latter. ``Superfluidity'' denotes
both the superconductivity of metals and the superfluidity of $^3$He. The
many-body hamiltonian of the system can be written as
\begin{equation}
H=H_{\rm pure}+\sum_jV_j\ .
\label{e.hamilton}\end{equation}
Here $H_{\rm pure}$ contains the kinetic energies of the particles and
interactions between them, and also interaction with the periodic crystal
lattice in the case of metals. It may also contain the interaction with
external magnetic field, for example. The latter term is the impurity
potential. In the case of metals, a common source for it is that some atoms of
the lattice are of different type than the others. In the case of
$^3$He, such a term is introduced by the aerogel. Quite generally, the
total impurity potential $V$ can be written as a sum so that each term $V_j$ is
nonzero only in some small region of space. These regions are referred to as
``scattering centers''.

The general quasiclassical theory for $H_{\rm pure}$ is discussed extensively
by Serene and Rainer (1983), and also elsewhere in this book. The
validity of this theory for the superfluid state relies on
the assumption $\lambda_{\rm f}\ll\xi_0$, which is well satisfied in $^3$He and
in most superconductors. The result is that instead of strongly interacting
particles, the system is better understood in terms of quasiparticles. The
quasiparticles propagate through the medium similar to classical particles. The
interactions between the quasiparticles are weak, and can be neglected in many
cases, which would not be the case for particles. Such quasiclassical
description is valid for properties that are dominated by processes taking
place near the Fermi surface in the momentum space. This includes virtually all
phenomena in the superfluid  state.

Let us consider adding the impurity effects to the quasiclassical theory. We
will not set a limitation to the magnitude of the impurity potential i.e.\ it
can be large, on the order of the Fermi energy $E_{\rm f}$. The crucial
assumption is that this large potential occurs only in a limited volume of the
total space. This means that the quasiparticle description is still valid in
most of the space. The average distance a quasiparticle can travel between it
collides with  scattering centers defines a mean free path $\ell$. In order
to use the quasiparticle picture, $\ell$ has to be large in comparison to
the Fermi wave length
$\lambda_{\rm f}$. In addition, we neglect coherent scattering from two or
more scattering centers. This is justified to leading order in
$\lambda_{\rm f}/\ell$ as long as the locations of the scattering centers can be
considered random on the scale of $\lambda_{\rm f}$. In other words, the
interference can be neglected in an ensemble average if for each scattering
center
$j$ we have a distribution function
$n_j({\bf r})$ ($\int d^3r\,n_j({\bf r})=1$) that is smooth on the scale of
$\lambda_{\rm f}$. Note that $n_j({\bf r})$ can still be a good approximation to
the delta function when looked at on the scale of $\xi_0$. In the case that all
scattering centers are identical, it is sufficient to specify only one
distribution
$n({\bf r})$, which is normalized to the total number $N$ of scattering
centers: $\int d^3r\,n({\bf r})=N$

In a small neighborhood of a scattering center, the state of the system can be
strongly different from the bulk. For example, liquid $^3$He may be in a
solid-like state because of the van der Waals interaction with an impurity. In
order to allow such changes, the treatment within a scattering center must be
fully quantum mechanical. However, such a calculation is not possible in
practice because even the wave function of a quasiparticle is not known in any
detail. Therefore, one has to introduce the properties of a scattering center by
some phenomenological parameters. A sufficient description is to know the
scattering $T$ matrix or, in case of an isotropic scattering center, the
scattering phase shifts $\delta_l$.  It is important to notice that these are
parameters that appear in the normal state, and they can in principle be
measured without need to go into the superfluid state. 

The discussion above was limited to the neighborhood of the Fermi surface,
where the quasiparticle picture is valid. Because of their strong potential,
the impurities also have an effect outside of this range. As a consequence  the
pairing interaction, for example, is modified by the impurities. However, this
effect is of short range because of the relatively high energy.
Therefore we can neglect such effects relying on the assumption
that the volume fraction of the scattering centers is small. Similar argument
gives that the Landau Fermi-liquid parameters ($F_i^{\rm s}$, $F_i^{\rm s}$), as
well as the density ($\rho$) and the dipole-dipole interaction constant
($g_{\rm d}$) in $^3$He are unchanged. 

The size of a scattering center has to be small compared to the length scale
one is interested to study. In the superfluid state this scale
is typically set by the coherence length $\xi_0$. Thus a scattering center has
to be small compared to
$\xi_0$, although it can be large compared to $\lambda_{\rm f}$. It should be
noted that the size of the scattering center does not limit the size of the
impurity. In order to consider a macroscopic body, we represent its surface by
a set of scattering centers. This is possible because for all practical bodies,
the height-height correlations $\langle \zeta({\bf r})\zeta({\bf r}+\delta{\bf
r})\rangle$ of the surface have a distribution that is much wider that
$\lambda_{\rm f}$ when the separation $\delta r$ is on
the order of $\xi_0$. Thus the scattering becomes incoherent on this scale, and
therefore can be represented by different   scattering centers. Assuming that
the particles cannot penetrate into the body, the volume of the body should be
excluded from the calculation. For example, we estimate the volume of scattering
centers in aerogel as the volume of few atomic layers on the strands
($\approx\lambda_{\rm f}A$) rather than the volume of the strands themselves
($V$). (It happens,  however, that both quantities are on the same order of
magnitude in this case). 

Let us comment the effect of the pairing state. Superfluidity arises because
part of the quasiparticles are weakly bound to pairs. The momenta of the
quasiparticles in a pair can be denoted by ${\bf p}$ and $-{\bf p}$ because the
total momentum of the pair is small. The dependence of the pair wave function on
the direction $\hat{\bf p}={\bf p}/p$ can be described using spherical harmonic
functions and classified as $s$, $p$, $d$, etc.\ states. In most superconducting
metals the Cooper pairs form dominantly in an $s$-wave state. In $^3$He the
pairs form in a $p$-wave state, and there is increasing evidence that $d$ waves
are dominant in high-$T_{\rm c}$ cuprate superconductors. The amplitude of the
superfluid state and the transition  temperature can be calculated for these
cases as a function of impurity scattering, as will be discussed in more detail
later. In the quasiclassical approximation one finds that the impurities have no
effect on a homogeneous $s$-wave superfluid. For other waves the superfluidity
is suppressed by impurity, and it disappears when the mean free path is reduced
to $\sim\xi_0$. The interpretation is that scattering changes the momenta of the
quasiparticles, but this has no effect in the s-wave case because the wave
function is independent of the momentum direction. For other waves the sign of
the wave function changes for certain scattering angles, which leads to
destructive interference. There is an effect for $s$ waves, too, if the
superfluid state is inhomogeneous. Here the scattering makes the quasiparticles
more localized, and therefore they see more the same order parameter. This
effect makes the superconductor less sensitive to magnetic field, i.e.\ causes
the change from type I to type II superconductivity, as mentioned in the
introduction. 

All the discussion above has been in weak-coupling limit. This means neglecting
all corrections that are proportional to the small ratio $\lambda_{\rm
f}/\xi_0$. This approximation is clearly inadequate for some properties of
$^3$He. For example, it gives that the B phase is always more stable than the A
phase, contrary to the phase diagram in Fig.\
\ref{f.tcpres}. There are some calculations of strong coupling
corrections in the pure $^3$He (\cite{SRrev}). It seems very difficult to take
these corrections into account in the impure case, and therefore we do not
consider them here.

\section{Homogeneous Scattering Model}
\label{s.hsm}
We do not know the precise distribution $n({\bf r})$ of scattering centers in
aerogel. Therefore we have to take a model for $n({\bf r})$. The simplest
possible one is the homogeneous scattering model (HSM). There one assumes that
$n({\bf r})=n$ is a constant. In other words, the
probability for a quasiparticle to be scattered is the same at all locations
${\bf r}$. This same approximation is commonly used to study impurities in
superconductors (\cite{Gorkov,AG}).

The mean free path in the HSM is given by $\ell=(n\sigma)^{-1}$. Here $\sigma$
is the scattering cross section of a scattering center. We
additionally assume that the medium is isotropic, i.e.\ 
$\ell$ is independent of the direction of quasiparticle momentum. We also
assume that the scattering is nonmagnetic. This means that the scattering
probability is the same for both directions of the spin, and the spin is not
changed in the scattering.  

A convenient property of the isotropic HSM is that both the Ginzburg-Landau
(GL) theory and Leggett's theory of NMR (\cite{Leggett}) have the same  form
as in pure $^3$He. The changes appear only via modified parameter values of
these theories. The GL theory allows a convenient way to represent
the results of the HSM at temperatures near $T_{\rm c}$. Therefore we describe
it in detail below. 

The order parameter in superfluid $^3$He is a complex $3\times 3$ matrix
$A_{\mu  i}$. This describes the wave function of the Cooper pairs such that
for a given direction
$\hat{\bf p}$ of momentum, the (unnormalized) spin wave function is
given by (\cite{LeggettRMP}) 
\begin{equation}
\psi_{\rm spin}(\hat{\bf
p})=(-d_x+id_y)\vert\uparrow\uparrow\rangle+d_z\vert
\uparrow\downarrow+\downarrow\uparrow\rangle+
(d_x+id_y)\vert\downarrow\downarrow\rangle\ ,
\label{e.cooperspin}
\end{equation}
where $d_\mu=\sum_iA_{\mu  i}\hat{p}_i$. This
apparently strange notation has the advantage that the spin $\mu$ and
the orbital $i$ indices in $A_{\mu  i}$ transform equally in coordinate
rotations. The amplitude of $A_{\mu  i}$ describes how strong the superfluid
state is. In particular, it goes to zero continuously when the temperature
approaches the superfluid transition temperature $T_{\rm c}$. 

The free energy $f$ of the superfluid is a function of the order
parameter $A_{\mu  i}$. Similar to the Ginzburg-Landau theory of
superconductivity, one can expand 
$f$ in powers of $A_{\mu  i}$ when the temperature is not much
different from $T_{\rm c}$.  The most important ``bulk'' terms in the expansion
are (\cite{MerminS}) 
\begin{eqnarray}
f_{\rm bulk}&=&f_{\rm n}+\alpha A_{\mu i}^*A_{\mu i}
+\beta_1\vert A_{\mu i}A_{\mu i}\vert^2
+\beta_2(A_{\mu i}A_{\mu i}^*)^2\nonumber \\\mbox{}&&
+\beta_3 A_{\mu i}^*A_{\nu i}^*A_{\nu j} A_{\mu j}
+\beta_4 A_{\mu i}^*A_{\nu i}A_{\nu j}^*A_{\mu j}
+\beta_5 A_{\mu i}^*A_{\nu i}A_{\nu j}A_{\mu j}^*\ .
\label{e.gl}
\end{eqnarray}
Here a summation over repeated indices is implied.
Equation (\ref{e.gl}) includes the allowed terms up to fourth order, since
several terms have to vanish by symmetry: $f$ has to be real valued, and it
should remain unchanged in rotations of both the spin and the orbital parts of
$A_{\mu  i}$. This is because the pairing interaction is unchanged in such
rotations. For stability the fourth order terms have to be positive definite, but
otherwise the coefficients
$\beta_i$ are arbitrary in a phenomenological approach.

Minimizing (\ref{e.gl}) one finds the normal state ($A_{\mu  i}=0$) when
$\alpha>0$. In the superfluid state ($\alpha<0$) there are several
possible minima. The most important are the ones corresponding to the A and B
phases. The order parameter in the B phase has the form
\begin{equation}
A_{\mu  j}=\exp(i\vartheta)\Delta_{\rm B} R_{\mu  j}\ ,
\label{e.bphasea}
\end{equation}
where $R_{\mu  i}$ is an arbitrary rotation matrix ($R_{\mu  i}R_{\mu 
j}=\delta_{ij}$), $\vartheta$ an arbitrary phase, and 
\begin{equation}
f_{\rm B}-f_{\rm n}={3\over 2}\alpha\Delta_{\rm B}^2=-{3\alpha^2\over
4(3\beta_{12}+\beta_{345})}\ .
\label{e.bphasef}
\end{equation}
We use the notation $\beta_{ij\ldots}=\beta_i+\beta_j+\ldots$. The A phase has
\begin{equation}
A_{\mu  j}=\Delta_{\rm A}\hat d_\mu(\hat m_j+i\hat n_j)\ ,
\label{e.aphasea}
\end{equation}
where $\hat{\bf d}$ is an arbitrary unit vector and $\hat{\bf m}$, $\hat{\bf
n}$, and $\hat{\bf l}=\hat{\bf m}\times\hat{\bf n}$ from an arbitrary
orthonormal triad. The energy and amplitude are 
\begin{equation}
f_{\rm A}-f_{\rm n}=\alpha\Delta_{\rm A}^2=-{\alpha^2\over
4\beta_{245}}\ .
\label{e.aphasef}
\end{equation}

In order to get the coefficients $\alpha$ and $\beta_i$ one has to solve
the quasiclassical equations. This is briefly explained in the
Appendix. The result for $\alpha$ is 
\begin{equation}
\alpha={N(0)\over 3}\left[\ln{T\over 
T_{\rm c0}}+\sum_{n=1}^\infty\left({1\over n-{1\over 2}}-{1\over
n-{1\over 2}+x}\right)\right]\ ,\label{e.alpha}
\end{equation}
where $2N(0)$ is the density of states at the Fermi surface and $x=\hbar v_{\rm
f}/4\pi T\ell_{\rm tr}$. The scattering parameter is the {\it transport}
mean free path $\ell_{\rm tr}=(n\sigma_{\rm tr})^{-1}$. The transport cross
section $\sigma_{\rm tr}$ gives more weight to large
scattering angles $\theta$ that the total cross section $\sigma$. The precise
definitions are 
$\sigma_{\rm tr}=
4\pi\langle(1-\cos\theta)(d\sigma/d\Omega)(\theta)\rangle_\Omega$ and
$\sigma=
4\pi\langle(d\sigma/d\Omega)(\theta)\rangle_\Omega$. Here 
$(d\sigma/d\Omega)(\theta)$ is the differential scattering cross
section and
$\langle\ldots\rangle_\Omega$ denotes average over $4\pi$ solid angle.

\begin{figure}[bt]
\begin{center}\leavevmode
\includegraphics[width=0.5\linewidth]{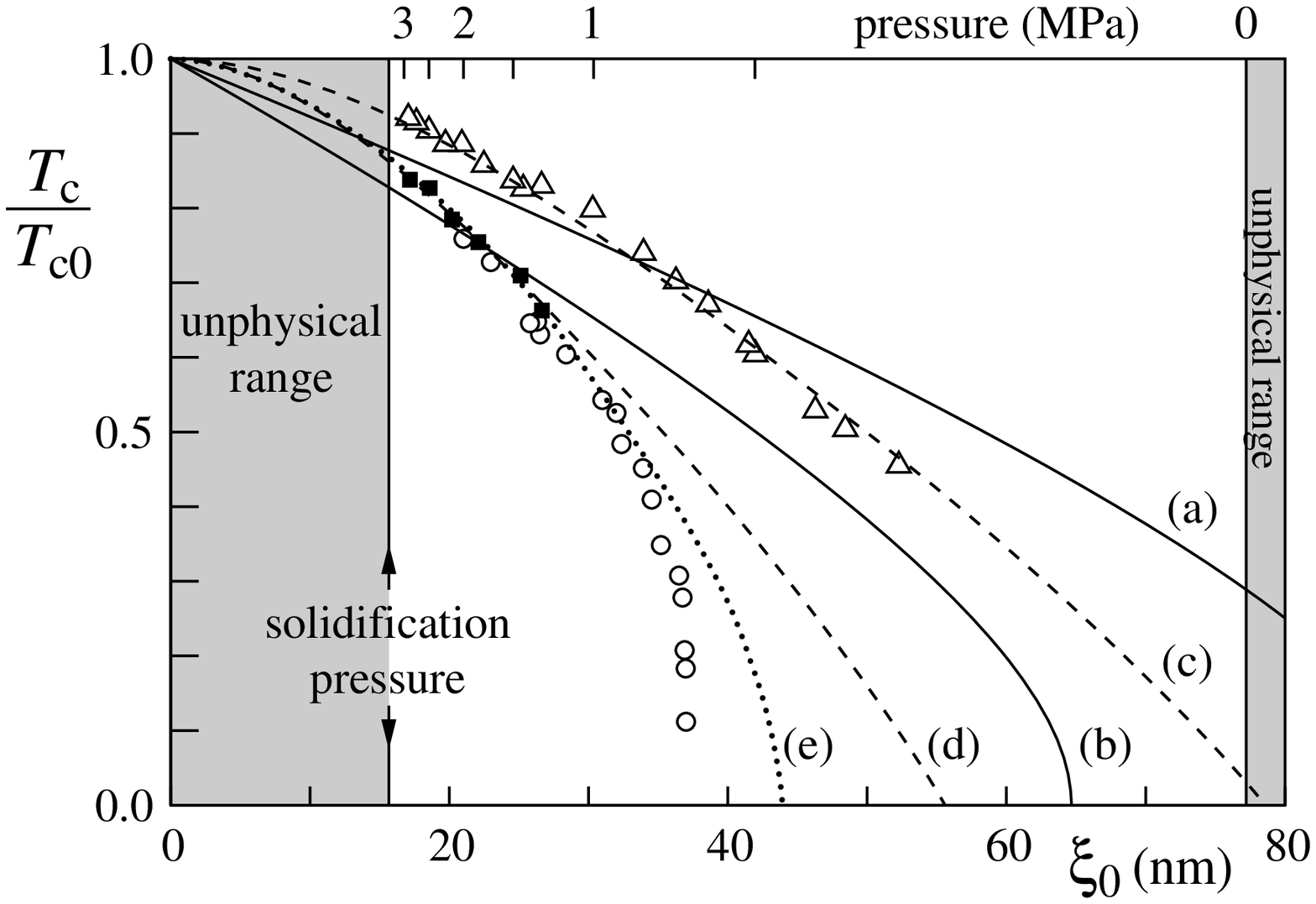}
\bigskip
\caption[f.tc]{ 
The transition temperature in aerogel relative to that in bulk,
$T_{\rm c}/T_{\rm c0}$. The horizontal axis is the coherence length
$\xi_0={\hbar v_{\rm f}/ 2\pi k_{\rm B}T_{\rm c0}}$. The experimentally
inaccessible values of $\xi_0$ are denoted by shading. The data points
are the same as in Fig.\ \ref{f.tcpres}:   
Porto and Parpia (1995, \font\apu=cmsy7  {\apu\char'064}),
Sprague et al. (1995, \kern .2ex\vrule height .95ex width .8ex depth -.15ex
\kern .2ex), and Matsumoto et al. (1997, $\circ$).
The solid lines correspond to the  homogeneous scattering model (HSM) at mean
free paths $\ell_{\rm tr}=320$ nm  (a) and  230 nm (b), and the dashed lines
correspond to slabs of thicknesses $D=105$ nm (c) and 74  nm (d). The dotted
line is the isotropic inhomogeneous scattering model (IISM) with average mean
free path 70 nm, unit cell radius $R =140 $\ nm, and scattering profile
parameter $j=8$. 
}\label{f.tc}\end{center}\end{figure}

The $\alpha$ term (\ref{e.alpha}) was first calculated by Larkin (1965). It has
exactly the same form as calculated earlier for {\it magnetic} impurities in
$s$-wave superfluid (\cite{AG}). The transition temperature 
$T_{\rm c}$ obtained from the condition $\alpha(T_{\rm c})=0$ is plotted in 
Fig.\ \ref{f.tc}.
The figure also shows the experimental data plotted as
a function of the coherence length $\xi_0$. We see that the HSM gives a
suppression of $T_{\rm c}$, but its dependence on $\xi_0$ is much weaker than
is seen experimentally. 

We have assumed in Fig.\ \ref{f.tc} that $\ell_{\rm tr}$ is a
pressure-independent constant. We can identify three effects that could
make $\ell_{\rm tr}=(n\sigma_{\rm tr})^{-1}$ pressure dependent. (i) The
structure of aerogel changes. (ii) The Fermi momentum changes. (iii) The wave
function of a quasiparticle changes.  It seems that all these
effects are negligible compared to the changes shown in Fig.\ \ref{f.tc}. 
In particular, the Fermi momentum $\hbar k_{\rm f}$ changes only 10\% over the
whole pressure range, and calculations with hard spheres indicate that
$\sigma_{\rm tr}$ is very weakly dependent on $k_{\rm f}$ for impurities that
are large compared to $\lambda_{\rm f}$. What remains most uncertain is point
(ii) because we do not know the wave function of a quasiparticle and how it
changes with pressure.

In the neighborhood of the transition temperature $T_{\rm c}$ we expand
$\alpha(T)=\alpha'\times(T/T_{\rm c}-1)$. The higher
order corrections in the Taylor series are dropped because they would not improve
the accuracy of the GL functional (\ref{e.gl}). From equation
(\ref{e.alpha}) one easily gets
\begin{equation} \alpha'={N(0)\over
3}\left[1-x_{\rm c}\sum_{n=1}^\infty(n-{\textstyle{1\over 2}}+x_{\rm
c})^{-2}\right]\ ,\label{e.alphap} \end{equation}
where $x_{\rm c}=\hbar v_{\rm f}/4\pi T_{\rm c}\ell_{\rm tr}$.

It is not easy to calculate the fourth order coefficients $\beta_i$ with the
same generality as $\alpha$. Therefore we make the additional assumption that
only the $s$-wave scattering phase shift $\delta_0$ is nonzero. In this case
$\sigma_{\rm tr}=\sigma$, and we get 
\begin{eqnarray} \left(\matrix{\beta_1\cr
\beta_2\cr \beta_3\cr \beta_4\cr \beta_5\cr}\right)&=&a\left(\matrix{-1/2\cr
1\cr 1\cr 1\cr -1\cr}\right)+b\left(\matrix{0\cr 1\cr 0\cr 1\cr -1\cr}\right)
+\left(\matrix{\Delta\beta_1^{\rm sc}\cr \Delta\beta_2^{\rm sc}\cr 
\Delta\beta_3^{\rm sc}\cr \Delta\beta_4^{\rm sc}\cr
\Delta\beta_5^{\rm sc}\cr}\right)
\label{e.beta}\\
a&=&{N(0)\over 120(\pi T_{\rm c})^2}\sum_{n=1}^\infty(n-{\textstyle{1\over
2}}+x_{\rm c})^{- 3}\nonumber \\
b&=&{N(0)\hbar v_{\rm f}\over 288(\pi T_{\rm c})^3\ell }\left(\sin^2\delta_0-
{\textstyle{1\over 2}}\right)
\sum_{n=1}^\infty(n-{\textstyle{1\over 2}}+x_{\rm c})^{-4}\ .\nonumber 
\end{eqnarray}
Here $\Delta\beta_j^{\rm sc}$ are strong-coupling
corrections which are not evaluated here. A model calculation for them in pure
$^3$He is given by Sauls and Serene (1980). 

Note that $\alpha$ and $a$, instead
of depending separately on
$n$ and the cross section
$\sigma=(4\pi/k_{\rm f}^2)\sin^2\delta_0$, depend only on the 
combination
$\ell=(\sigma  n)^{-1}$. This is not the case for $b$, which is also
a function of 
$\delta_0$. The limiting cases $\sin^2\delta_0\rightarrow 0$ and
$\sin^2\delta_0=1$ are known as Born
and unitarity limits, respectively. This additional degree of freedom decreases
the precision of the theoretical results because $\delta_0$ is not generally
known. However, we argue in the following that this degree of freedom is
averaged out in aerogel. Firstly, if the impurities are not identical, we can
expect that there is a distribution of $\delta_0$'s instead of a single value.
Secondly, the calculation including all partial waves has been done in the case
of B phase (\cite{TKR}). There one finds a similar spread of the results as a
function of the phase shifts. But for large impurities the phase shifts
$\delta_l$ with different $l$ are essentially random, as demonstrated by
$\delta_l$ for hard spheres. Both these arguments support that a reasonable
guess, if no more detailed information exists, is to assume random phase shifts
$\sin^2\delta_0\rightarrow 0.5$. 

A necessary condition for the stability of the A phase is $f_{\rm
A}<f_{\rm B}$. Using equations (\ref{e.bphasef}), (\ref{e.aphasef}), and
(\ref{e.beta}) this reduces to 
\begin{equation}
a< 6\Delta\beta_1^{\rm sc}+2\Delta\beta_3^{\rm sc}-4\Delta\beta_{45}^{\rm
sc}\ .\label{e.astability}
\end{equation}
Because $a$ increases monotonically by factor 3 with decreasing $T_{\rm c}$,
the B phase becomes more favored with increasing scattering assuming that
$\Delta\beta_j^{\rm sc}$ do not grow even more. Assuming $\Delta\beta_j^{\rm
sc}$ remain constants, we can expect stable A phase only at
temperatures (in mK) and pressures where it is stable in the pure case.
Calculations with other phases indicate that they are not serious competitors to
A and B phases.

In order to make more comparisons with experiments, we need to consider
additional terms in the GL functional
$F=\int d^3rf({\bf r})$. These are the gradient energy
\begin{equation}
f_{\rm k}=K\left[(\gamma-1)\partial_iA_{\mu i}\partial_jA_{\mu j}^*
+\partial_iA_{\mu j}\partial_iA_{\mu j}^*\right]\ ,
\label{e.GLgradient}
\end{equation}
the energy of the magnetic field ${\bf H}$,
\begin{equation}
f_{\rm z}=-{\textstyle{1\over
2}}\chi_{\rm n}^{-1}H^2+g_{\rm z}H_\mu A_{\mu i}A_{\nu i}^*H_\nu\ ,
\label{e.GLfield}
\end{equation}
where $\chi_{\rm n}$ is the susceptibility of the normal state, and the magnetic
dipole-dipole interaction energy  
\begin{equation}
f_{\rm d}=g_{\rm d}(\vert A_{i i}\vert^2+A_{i j}A_{j i}^*-{\textstyle{2\over
3}}A_{\mu i}A_{\mu i}^*)\ .
\label{e.GLdipole}
\end{equation}
Here $f_{\rm k}$ is purely phenomenological similar to the bulk energy
(\ref{e.gl}), but in $f_{\rm z}$ and 
$f_{\rm d}$ we have dropped some terms, which could be there in a pure
phenomenological theory. For the coefficients we calculate
(Appendix) 
\begin{eqnarray} K&=&{N(0)\hbar^2 v_{\rm f}^2\over 240\pi^2 T_{\rm
c}^2}\sum_{n=1}^\infty(n-{\textstyle{1\over 2}}+x_{\rm c})^{-3}
\label{e.k}\\
\gamma&=&3+{5\hbar v_{\rm f}\over 12\pi T_{\rm
c}\ell}{\sum_{n=1}^\infty(n-{1\over 2})^{- 1}(n-{1\over 2}+x_{\rm
c})^{-3}\over\sum_{n=1}^\infty(n-{1\over 2}+x_{\rm c})^{-3}}
\label{e.gamma}\\
g_{\rm z}&=&{\hbar^2\tilde\gamma^2N(0)\over 48\pi^2 
T_{\rm c}^2(1+F_0^{\rm a})^2}\sum_{n=1}^\infty(n-{\textstyle{1\over 2}}+x_{\rm
c})^{-3}
\label{e.gz}\\
g_{\rm d}&=&{\mu_0\over 40}
\left[\hbar\tilde\gamma N(0)R\sum_{n=1}^{\epsilon_{\rm cut\,off}/2\pi
T_{\rm c}}(n-{\textstyle{1\over 2}}+x_{\rm c})^{-1}\right]^2,
\label{e.gd}
\end{eqnarray}
where $\tilde\gamma$ is the gyromagnetic ratio, $F_0^{\rm a}$ a
Fermi-liquid parameter, $R$ a renormalization
constant for the dipole energy, and $\epsilon_{\rm cut\,off}$ a high-energy
cut-off (\cite{Leggett}). The dipole-dipole coupling constant $g_{\rm d}$ is
different form the other coefficients
[(\ref{e.alpha})-(\ref{e.beta}), (\ref{e.k})-(\ref{e.gz})] because it is
dominated by high-energy processes (it would diverge for
$\epsilon_{\rm cut\,off}\rightarrow\infty$), and therefore it is a constant
(independent of scattering). 

All the coefficients above reduce to the well know GL coefficients in the limit
vanishing scattering, $\ell\rightarrow\infty$ (Fetter 1975, \cite{T87}). 

In most cases the terms $f_{\rm z}$ and $f_{\rm d}$ are small compared to $f_{\rm
bulk}$. Treating them as perturbations, we find in the A phase (apart from
constants)
\begin{eqnarray}
f_{\rm zA}&=&-{\textstyle{1\over
2}}\chi_{\rm n}^{-1}H^2+g_{\rm z}\Delta_{\rm A}^2(\hat{\bf d}\cdot{\bf H})^2
\label{e.GLfieldA}
\\
f_{\rm dA}&=&-2g_{\rm d}\Delta_{\rm A}^2(\hat{\bf d}\cdot\hat{\bf l})^2\ ,
\label{e.GLdipoleA}
\end{eqnarray}
which favor $\hat{\bf d}\perp{\bf H}$ and $\hat{\bf
d}\parallel\hat{\bf l}$, respectively. The B phase is slightly more complicated.
We parameterize the rotation matrix
$R_{\mu i}$ by an angle
$\theta$ and an axis $\hat{\bf  n}$ of rotation. The dipole-dipole
energy $f_{\rm d}$ favors $\theta={\rm arccos}(-1/4)=104^\circ$. The external
field leads to distortion of B phase order parameter (\ref{e.bphasea}). This
gives rise to the energy term  
\begin{equation}
f_{\rm dzB}=-{5g_{\rm d}g_{\rm d}\over 4\beta_{345}}(\hat{\bf n}\cdot{\bf
H})^2\ ,
\label{e.dz}
\end{equation}
which favors $\hat{\bf n}\parallel{\bf H}$.

For comparison to experiments, we still need to relate the observables to the
GL coefficients. The torsional oscillator experiments measure the superfluid
density $\rho_{\rm s}$ (\cite{Porto,Matsumoto}). This can be calculated by
evaluating $f_{\rm k}$ (\ref{e.GLgradient}) in the presence of a phase
gradient:
${\mbox{\boldmath$\nabla$}} A_{\mu j}=i{\bf q}A_{\mu j}$, where the
superfluid velocity ${\bf v}_{\rm s}=\hbar {\bf q}/2m_3$ and $m_3$ is the mass
of a
$^3$He atom. The main quantities measured in the NMR experiment are the 
magnetic susceptibility
$\chi$ and the shift of the resonance frequency
$\delta\omega$ from the Larmor value. The former
is obtained by evaluating $f_{\rm z}$. The latter needs evaluating $f_{\rm d}$
in the dynamic case  
(\cite{Leggett}). The results for homogeneous A and B phases are
\begin{eqnarray}
\rho_{\rm sA}&=&{8m_3^2\over \hbar^2}K\Delta_{\rm
A}^2[\gamma+1-(\gamma-1)(\hat{\bf  l}\cdot\hat{\bf v}_{\rm s})^2]
\label{e.rhoA}\\
\rho_{\rm sB}&=&{8m_3^2\over \hbar^2}(\gamma+2)K\Delta_{\rm B}^2
\label{e.rhoB}\\
\chi_{\rm A}&=&\chi_{\rm n} \ \ \ \ \ \ \ \ \ \ \ \ \ \ \ \ \ \ \ \ \ \ \ \ \ \
\ \ 
\ \ \ \ \ \ \ \ \ \ \ \ \ \ \ \ \ \ \ \ \ \ (\hat{\bf d}\perp{\bf H})
\label{e.chiA}\\
\chi_{\rm B}&=&\chi_{\rm n}-2g_{\rm z}\Delta_{\rm B}^2
\label{e.chiB}\\
\delta\omega_{\rm A}&=&{2\tilde\gamma\over\chi_{\rm n} H} g_{\rm d}\Delta_{\rm
A}^2[(\hat{\bf  l}\cdot\hat{\bf d})^2-(\hat{\bf l}\cdot\hat{\bf H})^2]\ \ 
\ \ \ \ \ \ \ \ \ \ \ \ \ \ (\hat{\bf d}\perp{\bf H})
\label{e.omegaA}\\
\delta\omega_{\rm B}&=&{15\tilde\gamma\over 2\chi_{\rm n} H} g_{\rm
d}\Delta_{\rm B}^2\vert\hat{\bf  n}\times\hat{\bf H}\vert^2\ \ \ \ \ \ \ \ \ 
\ \ \ \ \ \ \ \ \ \ \ \ \ \ \ \ (\theta=104^\circ)\ .
\label{e.omegaB}
\end{eqnarray}
Here $\hat{\bf v}_{\rm s}$ and $\hat{\bf H}$ denote the directions of the
superfluid velocity and field, respectively. The
susceptibility of the normal phase
$\chi_{\rm n}$ has to include the inert layer of $^3$He atoms on the
aerogel strands, which in fact is the dominant contribution at low
temperatures (\cite{Sprague1}). The frequency shifts $\delta\omega$ are for
small tipping angles of the magnetization from the equilibrium direction
$\hat{\bf H}$, and the magnetic field $H$ is assumed
large in comparison to $\sqrt{g_{\rm d}/g_{\rm z}}\approx 0.2$ mT. The results
for
$\chi_{\rm A}$,
$\delta\omega_{\rm A}$, and
$\delta\omega_{\rm B}$ are limited to the conditions indicated in
parenthesis after each equation.

\begin{figure}[bt]
\begin{center}\leavevmode
\includegraphics[width=0.7\linewidth]{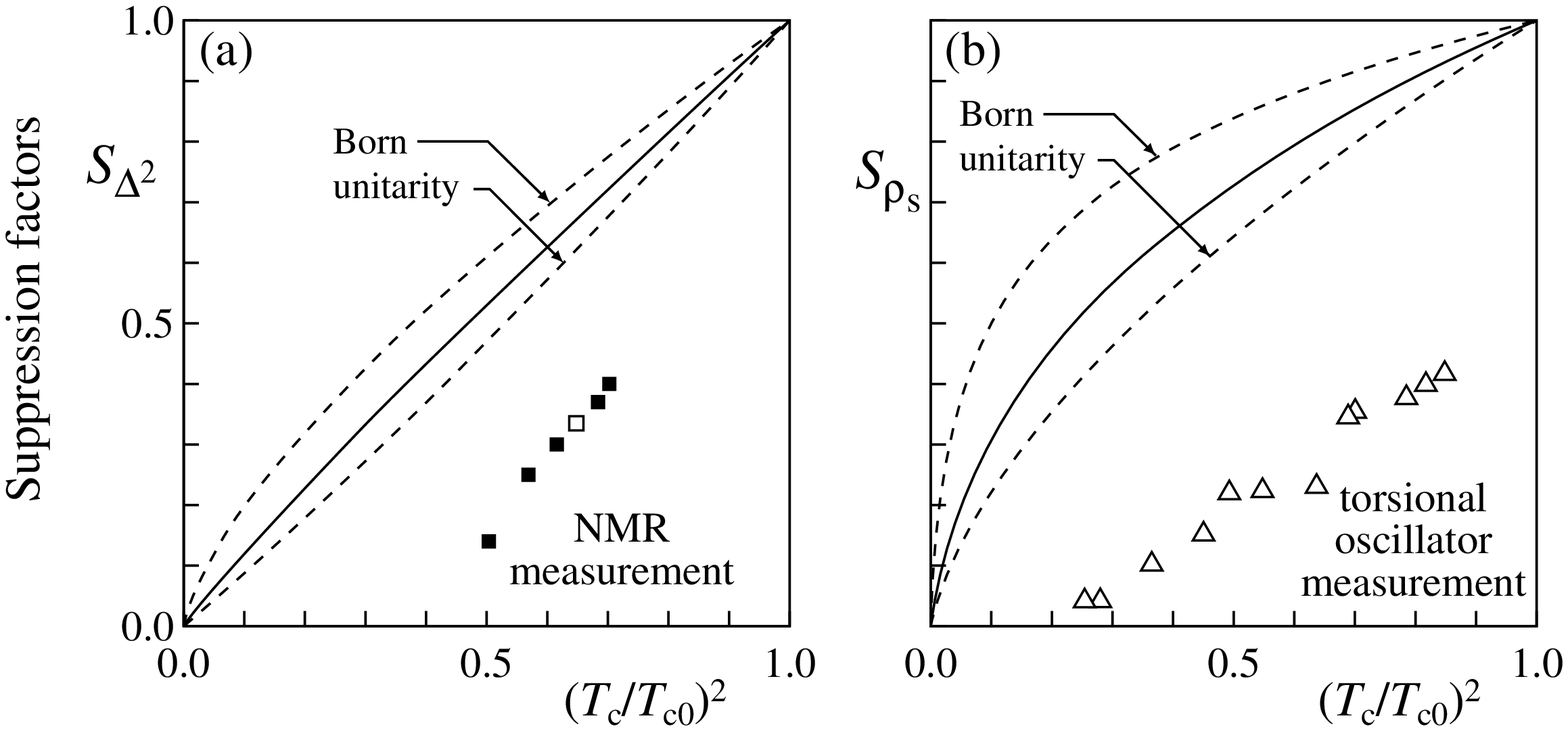}
\bigskip
\caption[f.sf]{ 
The suppression factors for $\Delta^2$ and $\rho_s$ as a function of $T_{\rm
c}$ reduction: $(T_{\rm
c}/T_{\rm
c0})^2$. The lines are theoretical results for the B phase when $t\rightarrow 1$
[see equations (\ref{e.sdelta}) and (\ref{e.srhos})]. The dashed lines denote
Born and unitarity limits, and solid lines are for the intermediate case
$\sin^2\delta_0=0.5$. The data points are from Sprague et al.
(1995, A type phase, \kern .2ex\vrule height .95ex width .8ex depth -.15ex
\kern .2ex), Sprague et al. (1996, B type
phase,\kern .2ex\vrule height .98ex width .1ex depth
-.18ex
\vrule height .28ex width .7ex depth -.18ex 
\kern -.8ex \vrule height .98ex width .7ex depth -.88ex
\vrule height .98ex width .1ex depth -.18ex\kern .2ex),
and Porto and Parpia (1995, \font\apu=cmsy7 
{\apu\char'064}), all corresponding to $t\approx 0.9$.
}\label{f.sf}\end{center}\end{figure}

The  NMR experiments with $^3$He (no $^4$He mixed) see no deviation of the
susceptibility from
$\chi_{\rm n}$ (\cite{Sprague1}, 1996). In the HSM model, we interpret
this as evidence for the A phase. Therefore, the measured frequency shift is
best compared with $\delta\omega_{\rm A}$. In Fig.\ \ref{f.sf}(a) we plot the
suppression factor 
\begin{equation}
S_{\chi_{\rm n}\delta\omega_{\rm A}}\equiv{\chi_{\rm n}\delta\omega(tT_{\rm
c})\over
\chi_{\rm n0}\delta\omega_{\rm A0}(tT_{\rm c0})}
= {\Delta^2(tT_{\rm c})\over
\Delta_0^2(tT_{\rm c0})}\equiv S_{\Delta^2}\ .
\label{e.sdelta}
\end{equation}
Generally, the suppression factor is the ratio of a quantity in
aerogel relative to the same quantity in the pure case. The latter is denoted
by subindex
$0$. The ratio is taken at the same temperature $t$ measured relative to the
corresponding $T_{\rm c}$. We consider here $t$ in the range $0.9\ldots 1$ where
the GL theory is expected to be valid. The middle equality  (\ref{e.sdelta}) is
based on equation (\ref{e.omegaA}). It is valid because
$g_{\rm d}$ is a constant and  the minimum-energy orientations of $\hat{\bf
d}$ and
$\hat{\bf l}$ are unchanged by scattering. (The latter is not necessarily
satisfied in generalizations of the HSM, see Section \ref{s.ahsm}.) Similarly we
can define a suppression factor for $\rho_{\rm s}$:
\begin{equation}
S_{\rho_{\rm s}}\equiv {\rho_{\rm s}(tT_{\rm c})\over
\rho_{\rm s0}(tT_{\rm c0})}\ .
\label{e.srhos}
\end{equation}
The B phase $\rho_{\rm s0}$ is used as a reference in the
experimental points of Fig.\ \ref{f.sf}(b).

The theoretical suppression factors $S_{\Delta^2}$ and $S_{\rho_{\rm s}}$ are
plotted in Fig.\ \ref{f.sf}. In these the impure B phase is compared with bulk
B phase. Very similar suppression factors are obtained when impure A phase is
compared with bulk A phase: The solid lines are the same for $\Delta^2$ and
also for $\rho_{\rm s}$ (\ref{e.rhoA}) averaged over all orientations of the A
phase [corresponding to $(\hat{\bf  l}\cdot\hat{\bf v}_{\rm
s})^2\rightarrow{1\over 3}$]. The variation between Born and unitarity limits
is slightly smaller in the A than in the B phase. 

We see from Fig.\ \ref{f.sf} that HSM indeed gives a suppression of $\Delta^2$
and $\rho_{\rm s}$, but it is insufficient to explain quantitatively the
measured suppression.

\section{Inhomogeneous scattering models}\label{s.iism}

The disagreement between the experiments and the HSM is clear in Fig.\
\ref{f.sf}.  We have checked that this failure does not arise from limitation
to the neighborhood of
$T_{\rm c}$ or to the $s$-wave scattering approximation. Also magnetic
scattering does not seem to give a solution to this problem. We also
discuss in Section \ref{s.ahsm} that the problem cannot be explained away by
random anisotropy of the aerogel. Therefore it seems inevitable to sacrifice to
basic assumption of the HSM: the homogeneity of the scattering. 

A simple model of inhomogeneous scattering is to consider $^3$He between two
diffusively scattering planes. The transition temperature for this ``slab
model'' is calculated by Kj\"aldman et al.\ (1978), and the suppression factors
are evaluated by Thuneberg et al.\ (1996). This gave promising results (see
Fig.\ \ref{f.tc}) but it is not a suitable model for aerogel because of its
strong anisotropy. Recently an  ``isotropic inhomogeneous scattering model''
(IISM) was studied (\cite{letter}). This is consistent with the observed
isotropy of aerogel, and it gives rather good fits to both $T_{\rm c}$ and
$S_{\Delta^2}$. Because only preliminary calculations on the IISM has been
done, we leave its discussion to another occasion. What is of
importance here that part of the HSM seems to remain valid:
Although the gap amplitude $\Delta^2$ is badly overestimated in
the HSM, the equations (\ref{e.rhoA})-(\ref{e.omegaB}) for $\rho_s$,
$\chi$, and $\delta\omega$ may still constitute a reasonable approximation. This
will be used in the following section. Similar arguments may be
applied to the HSM calculations by Baramidze et al. (1996).

\section{Anisotropic HSM}\label{s.ahsm}

In this section we consider how to generalize the isotropic HSM (Section
\ref{s.hsm}) to anisotropic scattering. The anisotropy will affect all the
terms in the GL functional, but the most important effect comes from the
modification of the second order bulk term (\ref{e.gl}). The anisotropy
requires the replacement 
\begin{equation}
\alpha A_{\mu i}^*A_{\mu i}\rightarrow\alpha_{ij} A_{\mu i}^*A_{\mu j}\ .
\label{e.replanis}
\end{equation}
So the scalar $\alpha$ is replaced by a tensor ${\underline\alpha}$. Solving
the quasiclassical equations in this case gives
\begin{equation}
{\underline\alpha}={N(0)\over 3}\left\{\ln{T\over 
T_{\rm c0}}+\sum_{m=1}^\infty\left[{1\over m-{1\over 2}}-\left(
m-{\textstyle{1\over 2}}+{\hbar v_{\rm
f}n\over 4\pi T}{\underline\sigma}_{\rm tr}\right)^{-1}\right]\right\}\ .
\label{e.alphaanis}
\end{equation}
Unit matrices multiplying $m-{1\over 2}$, for
example, are not shown explicitly. The transport cross-section tensor is defined
by the double angular average 
\begin{equation} {\underline\sigma}_{\rm tr}=\langle\,\hat{\bf
k}\left[\sigma(\hat{\bf k})\hat{\bf k} -4\pi\langle{d\sigma\over
d\Omega}(\hat{\bf k},\hat{\bf k}')\hat{\bf k}'\rangle_{\hat{\bf
k}'}\right]\rangle_{\hat{\bf k}}\ , \label{e.sigmaanis} \end{equation}
where $d\sigma/d\Omega$ is the differential scattering cross section and the
total cross section 
$\sigma(\hat{\bf k})=4\pi\langle (d\sigma/d\Omega)(\hat{\bf k},\hat{\bf k}')
\rangle_{\hat{\bf k}'}$. [In equation
(\ref{e.alphaanis}) we have assumed that both $\hat{\bf k}$ and $\hat{\bf k}'$
dependencies of $(d\sigma/d\Omega)(\hat{\bf k},\hat{\bf k}')$ can be represented
by $s$ and $p$-wave spherical harmonic functions.] For estimation of 
the anisotropy in aerogel we consider a long rod, which scatters
diffusely (randomly) the particles hitting it. This gives 
$n{\underline\sigma}_{\rm tr}=\ell^{-1}({9\over 8}-
{3\over 8}\hat{\bf a}\hat{\bf a})$, where $\hat{\bf a}$ is the direction of the
rod and $\ell$ denotes the average (transport) mean free path. Assuming the
anisotropy is small, the leading effect of the anisotropy can be represented
by
\begin{equation}
f_{\rm a}=-\alpha_1\hat a_iA_{\mu i}^*A_{\mu j}\hat a_j\ ,
\label{e.anis1}
\end{equation}
where 
\begin{equation}
\alpha_1={N(0)\hbar v_{\rm f}\over 32\pi
T_{\rm c}\ell}\sum_{m=1}^\infty(m-{\textstyle{1\over 2}}+x_{\rm c})^{-2}.
\label{e.alpha1}
\end{equation}
It is crucial that the anisotropy direction $\hat{\bf a}$ is not a constant.
We expect that the orientational correlation decays in a length $L_{\rm a}\sim
20$ nm, which is on the order of the average distance between strands. 

The anisotropy (\ref{e.anis1}) shifts the
transition temperature. It also changes the relative stability of different
phases. Instead of discussing these, we will here concentrate on how the
anisotropy can crucially modify the NMR properties. Some of the ideas below
are suggested by Volovik (1996).

Quite generally, we can write the following hydrodynamic free energy for the A
phase
\begin{eqnarray}
f_{\rm A}&=&-{\textstyle{1\over 2}}\lambda_{\rm d}(\hat{\bf d}\cdot\hat{\bf
l})^2+{\textstyle{1\over 2}}\lambda_{\rm z}(\hat{\bf d}\cdot{\bf
H})^2+{\textstyle{1\over 2}}\lambda_{\rm a}(\hat{\bf a}\cdot\hat{\bf
l})^2+{\textstyle{1\over 2}}\rho_\perp{\bf v}_{\rm s}^2+{\textstyle{1\over
2}}(\rho_\parallel-\rho_\perp)(\hat{\bf l}\cdot{\bf v}_{\rm s})^2 \nonumber
\\&&+C{\bf v}_{\rm s}\cdot\nabla\times\hat{\bf l} -C_0(\hat{\bf l}\cdot{\bf
v}_{\rm s}) (\hat{\bf l}\cdot\nabla\times\hat{\bf l})+
{\textstyle{1\over 2}}K_{\rm s}(\nabla\cdot\hat{\bf l})^2
+{\textstyle{1\over 2}}K_{\rm t}(\hat{\bf l}\cdot\nabla\times\hat{\bf
l})^2\nonumber
\\ &&+{\textstyle{1\over 2}}K_{\rm b}\vert\hat{\bf
l}\times(\nabla\times\hat{\bf l})\vert^2 +{\textstyle{1\over 2}}K_5
\vert(\hat{\bf l}\cdot\nabla)\hat{\bf d}\vert^2+ {\textstyle{1\over
2}}K_6[(\hat{\bf l}\times\nabla)_i\hat{\bf d}_j)]^2\ .
\label{e.f}
\end{eqnarray}
The first two terms are already familiar from equations (\ref{e.GLfieldA}) and 
(\ref{e.GLdipoleA}). The third is the anisotropy term (\ref{e.anis1}) with
$\lambda_{\rm a}=2\alpha_1\Delta_{\rm A}^2$. The rest arises from the
gradient energy (\ref{e.GLgradient}) when the A-phase  order parameter
(\ref{e.aphasea}) is substituted into it (\cite{Cross}). All the coefficients
are proportional to $\Delta_{\rm A}^2$. Because this common factor drops out in
relative comparisons of the terms, our estimations below are independent of
$\Delta_{\rm A}^2$.

The idea is that the random field $\hat{\bf a}$ tries to orient the 
$\hat{\bf l}$ vector. However, the gradient energy strongly limits the
variation of
$\hat{\bf l}$ on the scale $L_{\rm a}$. Instead, 
$\hat{\bf l}$ varies only on a ``orbital'' scale $L_{\rm o}\gg L_{\rm a}$.
In addition, the $\hat{\bf d}$ vector varies on a ``spin'' scale $L_{\rm
s}$, which also is large in comparison to ``aerogel'' scale $L_{\rm a}$.
We estimate $L_{\rm o}$ and $L_{\rm s}$ using arguments
presented by Imry and Ma (1975). The exact functional (\ref{e.f}) is
approximated by 
\begin{eqnarray}
f_{\rm A}&\approx&-{\lambda_{\rm d}\over 6}-{\eta\lambda_{\rm d}\over
3\eta+2[({L_{\rm s}/L_{\rm o}})^{3/4}-1]^2}
\nonumber\\&&\hbox{}
+{\lambda_{\rm
a}\over 6}-{\eta\lambda_{\rm a}\over 2}\left({L_{\rm a}\over L_{\rm
o}}\right)^{3/2}+ {K_{\rm s}\over 2L_{\rm s}^2}+{K_{\rm o}\over 2L_{\rm
o}^2}\ .\label{e.rfm}
\end{eqnarray} 
This is a functional of only two variables, $L_{\rm s}$ and $L_{\rm o}$. The
first two
terms constitute an interpolation of the dipole-dipole energy between two
limits. The minimum energy 
$-{1\over 2}\lambda_{\rm d}$ is obtained in the limit $L_{\rm s}=L_{\rm o}$.
In the opposite limit $L_{\rm s}\gg L_{\rm o}$, the dipole-dipole energy
consists of a random average $-{1\over 6}\lambda_{\rm d}$ plus a fluctuation
$-{1\over 2}\eta\lambda_{\rm d}(L_{\rm o}/L_{\rm s})^{3/2}$ (\cite{IM}). The
origin of the factor $(L_{\rm o}/L_{\rm s})^{3/2}$ is that in a box of volume
$L_{\rm s}^3$ there are $N=(L_{\rm o}/L_{\rm s})^3$ uncorrelated regions of
$\hat{\bf l}$, and therefore the total fluctuation they produce is
proportional $\sqrt{N}$. Here $\eta=2/3\sqrt{5}$ is the standard deviation of
$(\hat{\bf n}\cdot\hat{\bf z})^2$ for a random unit vector $\hat{\bf n}$. The
third and fourth terms in (\ref{e.rfm}) approximate similarly the anisotropy
energy ${1\over 2}\lambda_{\rm a}(\hat{\bf a}\cdot\hat{\bf
l})^2$, but there only the limiting form
$L_{\rm o}\gg L_{\rm a}$ is needed. The last two terms are simplified gradient
energies of
$\hat{\bf d}$ and
$\hat{\bf l}$. We assume
$K_{\rm s}\approx 12K\Delta_{\rm A}^2$ and $K_{\rm o}\approx 6K\Delta_{\rm
A}^2$, where a factor 3 arises from the dimensionality of the space.

The simplified energy functional can now be trivially minimized numerically,
and also to a great extent analytically. We find two basically
different solutions, which are described below. These solutions have equal
energy if $L_{\rm a}=48$ nm, which is only slightly larger than the estimated
distance between stands 20 nm. The numerical estimates use this and the
pressure of 2.8 MPa.

In the ``dipole locked'' solution $\hat{\bf d}$ and $\hat{\bf l}$ follow
closely each other, $\hat{\bf d}({\bf r})\approx\hat{\bf l}({\bf r})$. We find 
rather long $L_{\rm s}\approx L_{\rm o}\approx 70\ \mu$m.   The NMR
experiments are done in large magnetic field, so that  $\hat{\bf
d}\approx\hat{\bf l}$ is allowed to vary only in the plane perpendicular to
${\bf H}$. We see from equation (\ref{e.omegaA}) that the NMR frequency shift
$\delta\omega_{\rm A}$ is essentially unchanged from the HSM prediction. This
state has lowest energy for $L_{\rm a}<48$ nm.

In the ``dipole unlocked'' solution $\hat{\bf l}$ varies on a much shorter
scale than $\hat{\bf d}$, $L_{\rm o}\ll L_{\rm s}$. We find $L_{\rm o}\approx
8\ \mu$m and $L_{\rm s}\approx 3$ mm. Again the magnetic field limits $\hat{\bf
d}$ to the plane perpendicular to
${\bf H}$. However, $\hat{\bf l}$ is free to vary in all directions. The NMR
frequency shift is totally suppressed: it is reduced by factor $10^{-5}$
relative to the HSM. This is because $\langle(\hat{\bf  l}\cdot\hat{\bf
d})^2\rangle\approx\langle(\hat{\bf l}\cdot\hat{\bf H})^2\rangle\approx
{1\over 3}$ in the expression for $\delta\omega_{\rm A}$ (\ref{e.omegaA}).
This state has lowest energy if $L_{\rm a}>48$ nm.

As discussed above, the experiments see that the NMR frequency shift is
smaller than the HSM prediction. However, they still are on the same
order of magnitude. This strongly supports the dipole-locked state because in
the unlocked state the suppression would be much more severe. Moreover,
because the suppression in the locked state is unchanged from HSM, we reach the
conclusion (stated in Section \ref{s.iism}) that random anisotropy cannot
explain the difference between measurements and the HSM. 

The presence of the dipole-locked state of $^3$He in aerogel implies that the
anisotropy of aerogel is very small: it is equivalent to rods whose
directions are correlated over distance $L_{\rm a}<48$ nm. This invalidates the
slab model, because the thickness $D$ of the slab is larger that this (see Fig.\
\ref{f.tc}). 

Above we have considered the NMR properties only in the case that the
``tipping angle'' $\beta$ between the field and the magnetization is small.
When the angle is increased, the frequency shift $\delta\omega$ gradually
diminishes. This is similar to bulk $^3$He-A, and it is well understood
theoretically (\cite{BS,Fomin}). However, the frequency shift suddenly
disappears at about $\beta=40^\circ$, and remains zero for all larger angles
(\cite{Sprague1}). As suggested by Volovik (1996), this striking observation
can be associated with the locked$\rightarrow$unlocked transition discussed
above. At the time of writing this, this is still under study, so we postpone
the discussion to another occasion. 

\section{Conclusion}

We have extensively studied the homogeneous scattering model in the
Ginzburg-Landau approximation. Although it fails to produce the correct gap
amplitude, it may successfully applied in the hydrodynamic region. As an
example we considered NMR in the A phase. In the future it could be
used, e.g., for studying vortices of $^3$He in aerogel. 

\section*{Acknowledgments}
I thank my collaborators M. Fogelstr\"om, S.K. Yip, J.A. Sauls, R.
H\"anninen, and T. Set\"al\"a for various contributions to this work. Fruitful
discussions with W. Halperin, J. Hook, J. Parpia, J. Porto, D. Rainer, D.
Sprague, G. Kharadze, and G. Volovik are acknowledged.

\section*{Appendix}

In this appendix we briefly explain how the Ginzburg-Landau coefficients
($\alpha$,
$\beta_i$, etc.) are calculated in the quasiclassical theory.  
An intermediate quantity in the calculation is the quasiclassical
$4\times 4$ matrix Green's function $\hat g(\hat{\bf k},{\bf
r},\epsilon_m)$. The arguments are the direction of the momentum
$\hat{\bf k}$, the location ${\bf r}$, and the Matsubara frequencies
$\epsilon_m=\pi k_{\rm B}T(2m+1)$. The Green's
function is determined from the Eilenberger equations
\begin{eqnarray} &[i\epsilon_m\hat\tau_3-\hat\nu-\hat\rho- \hat\Delta, \hat
g]+iv_{\rm F}\hat{\bf k}\cdot\mbox{\boldmath$\nabla$}_{\bf r}\hat g=0& 
\label{e.eil}\\
& \hat g\hat g=-\pi^2\ .&
\label{e.norm}\end{eqnarray} 
Here $\hat\tau_i$
denote the Pauli matrices in Nambu space, and $[A,B]=AB-BA$.
For more details the reader is referred to the review article by Serene and
Rainer (1983). 
The self-consistency equations for the diagonal $\hat\nu$ and off-diagonal
$\hat\Delta$ self energies are given in
formulas (5.10) of Serene and Rainer (1983). The impurity self-energy
$\hat\rho(\hat{\bf k},{\bf r},\epsilon_m)$ equals $n({\bf r})\hat t(\hat{\bf
k},\hat{\bf k},{\bf r},\epsilon_m)$, where $n({\bf r})$ is the concentration
of the scattering centers. The $\hat t$-matrix of a single scattering center
is determined by
\begin{eqnarray} \hat t(\hat{\bf k},\hat{\bf k}',{\bf r},\epsilon_m)&=&\hat
v(\hat{\bf k},\hat{\bf k}')+N(0)\langle\hat v(\hat{\bf
k},\hat{\bf k}'')
\hat g(\hat{\bf k}'',{\bf
r},\epsilon_m)\hat t(\hat{\bf k}'',\hat{\bf k}',{\bf
r},\epsilon_m)\rangle_{\hat{\bf k}''}\ .\label{e.t}
\end{eqnarray}
Also we need the free energy functional which is given by equation (5.11) of
Serene and Rainer (1983).

The coefficients are calculated by solving Green's
function perturbatively $\hat g=\hat g_0+\hat g_1+\hat g_2+ \ldots$. Here  
$\hat g_0$ is the normal state Green's function, and the
subindex denotes order in $\hat\Delta$. Then one collects terms of order 0, 1,
2, and 3 in the Eileberger, self-consistency and
$\hat t$-matrix equations, and solves them in each order. Substitution to
energy functional gives the results $\alpha$ (\ref{e.alpha}) and
$\beta_i$ (\ref{e.beta}). For other coefficients it is sufficient to limit to
first order in
$\hat\Delta$, but one has to do additional expansions in gradients and in
magnetic field. Substitution to energy functional gives $K$ (\ref{e.k}),
$\gamma$ (\ref{e.gamma}) and $g_{\rm z}$ (\ref{e.gz}). The dipole
coefficient (\ref{e.gd}) is obtained using the dipole-dipole Hamiltonian
(\cite{Leggett}) in the general energy functional (5.6) of Serene and Rainer
(1983).

\end{document}